\documentclass[aps,prb,twocolumn,showpacs,twoside,10pt]{revtex4-1}
\usepackage[utf8]{inputenc}
\usepackage[english]{babel}
\usepackage{amssymb,amsmath}
\usepackage{graphicx}
\usepackage{color}
\usepackage[colorlinks,bookmarks=false,citecolor=blue,linkcolor=red,urlcolor=blue]{hyperref}
\usepackage{bbm}

\begin{document}
\title{Quantum-disordered ground state for hard-core bosons on the frustrated square lattice}
\author{Ansgar Kalz}
\email{kalz@theorie.physik.uni-goettingen.de}
\affiliation{Institut für Theoretische Physik, Universität Göttingen, 37077 Göttingen, Germany}
\author{Andreas Honecker}
\affiliation{Institut für Theoretische Physik, Universität Göttingen, 37077 Göttingen, Germany}
\author{Sebastian Fuchs}
\affiliation{Institut für Theoretische Physik, Universität Göttingen, 37077 Göttingen, Germany}
\author{Thomas Pruschke}
\affiliation{Institut für Theoretische Physik, Universität Göttingen, 37077 Göttingen, Germany}

%\keywords{hard-core bosons, frustrated lattice, ground-state phase diagram, QMC simulation}

\date{\today} 

\begin{abstract} 
We investigate the phase diagram of hard-core bosons on a square lattice with competing interactions. The hard-core bosons can also be represented by spin-$1/2$ operators and the model can therefore be mapped onto an anisotropic $J_1$-$J_2$-Heisenberg model. We find the N\'eel state and a collinear antiferromagnetic state as classical ordered phases to be suppressed by the introduction of ferromagnetic exchange terms in the $x$-$y$ plane which result in a ferromagnetic phase for large interactions. For an intermediate regime, the emergence of new quantum states like valence bond crystals or super-solids is predicted for similar models. We do not observe any signal for long-range order in terms of conventional order or dimer correlations in our model and find an exponential decay in the spin correlations. Hence, all evidence is pointing towards a quantum-disordered ground state for a small region in the phase diagram.  
\end{abstract}


\maketitle

\section{Introduction}
The investigation of frustrated spin models has become a rather active field over the past years due to the rising interest in new quantum phases like valence bond solids or spin liquids.\cite{B:misguich2,B:mila11} Another motivation to study these systems is the question of the microscopic origin of high-$T_C$ superconductivity which is still controversally discussed and often connected with frustrating spin interactions. One of the most interesting and challenging problems in this field is the $J_1$-$J_2$-spin-$1/2$ isotropic Heisenberg model.\cite{P:capriotti03, B:richter04, B:misguich2, P:mambrini06, P:darradi08, P:ralko09, P:richter10, P:richter10b} There is still no final answer to the question of the intermediate phase in the ground state phase diagram of this model and several techniques have been used to track it down. Non-variational Quantum Monte-Carlo (QMC) simulations have a severe sign problem for the frustrated model and are, hence, very limited for such a system. In this paper we present our work on an analogous bosonic model which maps onto the Heisenberg model \cite{P:batrouni00, P:batrouni01, P:chen08, P:melko08} for a certain set of parameters and may give some crucial hints for the completely frustrated model.  

Starting from the classical model without quantum fluctuations, which was analyzed in the early eighties by Landau and Binder \cite{P:landau80, P:lanbin80, P:lanbin85} and was further investigated in recent years,\cite{P:lopez93, P:lopez94, P:malakis06, P:kalz08, P:kalz09} we examine the quantum model for finite temperatures and extrapolate our QMC results to $T=0$ to draw a ground state phase diagram.

For the equivalent anisotropic Heisenberg model we find two classical magnetically ordered phases (N\'eel and collinear state) as ground states for small quantum fluctuations and a direct transition between these two antiferromagnetic configurations. For large fluctuations the system becomes ferromagnetic in the $x$-$y$-plane. Close to the highly frustrated point which is accompanied by a large ground-state degeneracy in the classical limit we find a region with no finite order parameter and interpret this state as quantum-disordered. In the bosonic language the antiferromagnetic states are described by boson-density waves with wave vectors $\vec q = (\pi,\pi)$ and $\vec q = (0,\pi)$ or $(\pi,0)$ respectively. The ferromagnetic in-plane order is interpreted as Bose condensation of the magnons and, hence, corresponds to a superfluid order in the bosonic model.\cite{P:bloch31, P:hoehler50, P:matsubara56, P:liu72} From now on we will use these terms exchangeably.

The paper is structured as followed: in the subsequent part we introduce the model and give an overview of the underlying physics and critical points of the system. In the third section we discuss some insights on the model via perturbation theory. The fourth section is divided into three subsections and dedicated to the QMC simulations. In the first part we briefly introduce the QMC method we used and explain the difficulties with the thermalization process for the frustrated model. Thereafter we present and discuss our observations of the magnetic and quantum-correlated observables that yield the phase diagrams at finite and zero temperature. In section five we show results from an exact diagonalization for one set of parameters. In the concluding part of the paper we will discuss our results and give an outlook for further calculations.

\section{Model}
The Hamiltonian of the model is given by summation over all interactions of nearest neighbors (NN) and next-nearest neighbors (NNN) and the according exchange integrals for these bonds:
\begin{align}
H_{\text{boson}} =&~ t_1 \sum_{\text{NN}} \left ( b^{\dagger}_i b^{\phantom{\dagger}}_j + \text{h.c.} \right)
+ V_1 \sum_{\text{NN}} n_i n_j \nonumber\\
+&~ t_2 \sum_{\text{NNN}} \left ( b^{\dagger}_i b^{\phantom{\dagger}}_j + \text{h.c.} \right)
+ V_2 \sum_{\text{NNN}} n_i n_j\,.\label{e:hamil}
\end{align}
The $b_i^{(\dagger)}$ are bosonic creation (annihilation) operators with $(b_i^{(\dagger)})^2 = 0$ (hard-core bosons) and $n_i = b_i^{\dagger} b_i^{\phantom{\dagger}}$ is the occupation number for the site $i$ (limited to $0$ or $1$). The model resides on a $N= L\times L$ square lattice with periodic boundary conditions (for a sketch see Fig.~\ref{f:local} below). The $V_i>0$ are chosen to be repulsive and the hopping integrals $t_i<0$ are negative.
Thus, mapping the bosonic operators onto spin-$1/2$ operators, the model is equivalent to the anisotropic $J_1$-$J_2$-spin-$1/2$ Heisenberg model
\begin{align}
H_{\text{HM}} =&~ J_1^{x,y}/2 \sum_{\text{NN}} \left ( S_i^+ S_j^- + \text{h.c.} \right)
+ J_1^z \sum_{\text{NN}} S_i^z S_j^z \nonumber\\
+&~ J_2^{x,y}/2 \sum_{\text{NNN}} \left ( S_i^+ S_j^- + \text{h.c.} \right)
+ J_2^z \sum_{\text{NNN}} S_i^z S_j^z\label{e:heisenberg}
\end{align}
using $t_i = J_i^{x,y}/2$ and $V =J_i^{z}$. This model is frustrated in the $S^z$ component (antiferromagnetic interactions, $J_i^z>0$) and not frustrated in the $S^x$ and $S^y$ components (ferromagnetic interactions, $J_i^{x,y}<0$). 
For $2t_i = J_i^{x,y} = J_i^z = V_i$ the isotropic Heisenberg model is recovered. Since we are only interested in the case of a half-filled bosonic model ($N/2$ bosons for the whole lattice) we do not take into account a chemical potential in \eqref{e:hamil} and, hence, the magnetic field in the corresponding antiferromagnetic Heisenberg model \eqref{e:heisenberg} is zero and we are working in the subspace of $\langle S^z \rangle = 0$. Earlier works calculating the bosonic model with $t_i <0$ -- as we do here -- have focused on the effect of a varying chemical potential.\cite{P:batrouni00, P:batrouni01, P:chen08, P:melko08} Furthermore, Refs.~\onlinecite{P:batrouni00, P:batrouni01,P:chen08} did not consider next-nearest neighbor hopping.

In the classical limit, i.\,e., for $t_i \rightarrow 0$, the Hamiltonian represents the antiferromagnetic $J_1$-$J_2$-Ising model which is highly frustrated in the region $J_2 \approx J_1/2$.\cite{P:landau80, P:lanbin85, B:lanbin00, P:kalz08, P:kalz09} The phase diagram contains two magnetically ordered ground states -- N\'eel order for $J_2 < J_1/2$ and collinear order for $J_2 > J_1/2$ -- and the paramagnetic phase for high temperatures. At $J_2 = J_1/2$ the ground state is degenerate (of order $2^{L+1}$) which leads to a suppression of the critical temperature and freezing problems in simple Monte-Carlo simulations. For small quantum fluctuations $|t_i|$ we expect the classical ordered phases to survive for low temperatures and to build the ground state. At the frustrated point $V_2=V_1/2$ we calculated second order perturbation terms by means of degenerate perturbation theory (DPT) to classify the behavior in the vicinity of the critical point (see part \ref{p:DPT}). For large $|t_i|$ we expect a superfluid phase which corresponds to long-range ferromagnetic order in the $x$-$y$ plane in the spin model. 

For the most interesting region (intermediate $|t_i|$ and $V_2 \approx V_1 /2)$ the emergence of new quantum states was predicted by Balents {\it et al.} \cite{P:bartosch05} for similar models. The smallest building block for these quantum states is a dimerized configuration of two spins which in our anisotropic model needs a short explanation. In the basis of total spin $S^{tot}$ of these two spins there exist two entangled states with $\langle S^z \rangle = 0$ given by:
\begin{align}
|S_0\rangle = | S^{tot}=0, S^z &= 0 \rangle = \frac{1}{\sqrt{2}}\left(| \uparrow \downarrow \rangle - | \downarrow \uparrow \rangle\right)\,,\nonumber\\
|S_1\rangle = | S^{tot}=1, S^z &= 0 \rangle = \frac{1}{\sqrt{2}}\left(| \uparrow \downarrow \rangle + | \downarrow \uparrow \rangle\right)\,.
\end{align}
The energies for these two eigenstates depend explicitly on the amplitude $t$ of the quantum fluctuations:
\begin{align}
&(t (S_i^+S_j^-+h.c.)+VS_i^zS_j^z) | S_0 \rangle = (-t-V/4) | S_0 \rangle\nonumber\,,\\
&(t (S_i^+S_j^-+h.c.)+VS_i^zS_j^z)| S_1 \rangle = (t-V/4) | S_1 \rangle \,.
\end{align}
Hence, for our anisotropic model with $t_i < 0$, the lowest energy is given by the $S_1$-dimer state.
The covering of the whole lattice with dimers allows for different ordered configurations as, e.\,g., a columnar or staggered valence bond solid or for non-static arrangements as a resonating valence bond solid. Apart from these dimerized states a superposition of the classical ordered states -- order in $S^z$ and in-plane order at the same time -- is feasible. In the bosonic language these states are called supersolids.\cite{P:liu72, P:wessel05, P:damle05, P:melko05, P:chen08, P:melko08}

\begin{figure}
\includegraphics[width=0.48\textwidth]{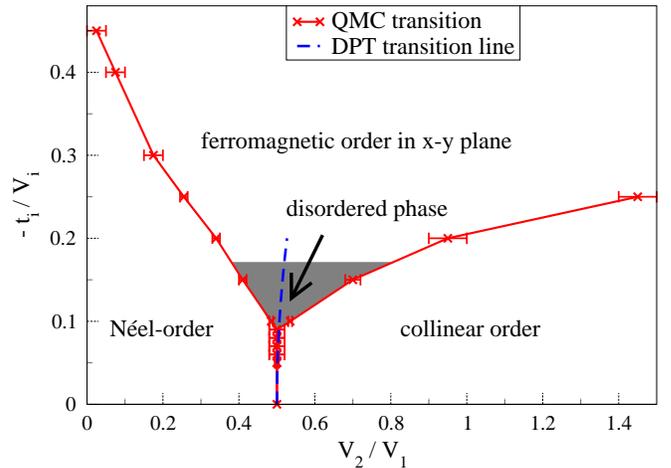}
\caption{\label{f:phaseT0}(Color online) Ground state phase diagram calculated with QMC simulations. As abscissa we plotted the degree of frustration $V_2/V_1$ and as ordinate the magnitude of the quantum fluctuations $-t_i/V_i$ (with $t_1/V_1 = t_2/V_2$). The dashed blue line indicates the direct transition from N\'eel to collinear order derived by means of DPT (up to second order, see section \ref{p:DPT}). The gray area represents roughly the region where we do not find any finite signal for various order parameters.}
\end{figure}

\begin{figure}
\includegraphics[width=0.48\textwidth]{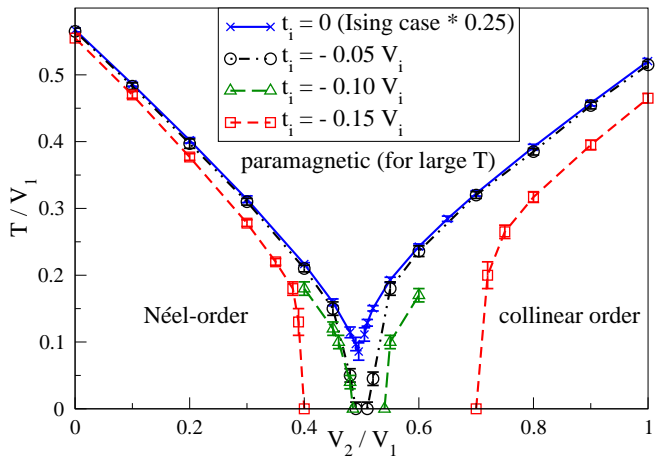}
\caption{\label{f:phaseT}(Color online) Phase diagrams at finite temperatures for different values of $t_i$. For increasing $|t_i|$ the transition temperatures are shifted to lower values and conventional order is suppressed. As a guide to the eye the full (blue) line shows the known phase boundary for the classical appropiately scaled Ising model.\cite{P:kalz08} For $t_i = -0.1~V_i$ (broadly dashed (green) line) only transition temperatures close to the critical point were calculated.}
\end{figure}

As the main result which will be derived below we already show the ground state phase diagram in Fig.~\ref{f:phaseT0}
where the suppression of the two magnetically ordered phases can be seen. Furthermore, it can be seen that the onset of the ferromagnetic phase for large fluctuations does not coincide with the opening of the two antiferromagnetic phases. Hence, we find an intermediate phase without conventional order. Furthermore, we present the finite-temperature phase diagrams for different magnitudes of quantum fluctuations in Fig.~\ref{f:phaseT}.
It can be seen that the classical ordering process is suppressed to lower temperatures due to the kinetic energy introduced by the exchange integrals $t_i$.

\section{Degenerate Perturbation Theory \label{p:DPT}}
To estimate the influence of small quantum fluctuations on the classical ground states nearby the critical point we calculated the second-order perturbation in the degenerate ground state manifold at the critical point.\cite{P:suzuki83} For $V_2=V_1/2$ every state with $2$ bosons per plaquette (square of $4$ lattice sites) has the same classical energy. Thus, for the whole lattice which is constituted of overlapping plaquettes this local degeneracy yields a global one of the order of the lattice length $L$, as explained in reference \onlinecite{P:kalz08}.
The DPT distinguishes between diagonal perturbations which leave the system in exactly the same state and off-diagonal perturbations which transfer the system into another state of the degenerate manifold. In the case of the frustrated square lattice different ground states are connected via flips of antiferromagnetic spins in a \emph{whole line or row} of the lattice. Thus, the order of off-diagonal perturbations scales with the length of the lattice $L$. Off-diagonal perturbations are therefore negligible. However, diagonal perturbations are already relevant in second order and non-constant for the two classical starting points -- N\'eel and collinear configurations. In the N\'eel state only nearest-neighbor hopping $t_1$ is possible on $2~L^2$ bonds of the lattice and yields an energy gain of $\Delta E_1 = -2~L^2\frac{t_1^2}{V_1}$. In the collinear state nearest-neighbor hopping on $L^2$ bonds and next-nearest-neighbor hopping on $2~L^2$ bonds is possible which gives an energy gain $\Delta E_2 = -2~L^2\frac{t_2^2}{3V_2}-L^2\frac{t_1^2}{V_1}$. Thus, small fluctuations enforce the classical ground states in the vicinity of the critical point. Calculating the transition line between the N\'eel and collinear state -- taking into account only second order corrections to the classical energies and using  $t_2\approx t_1/2$ and $V_2\approx V_1/2$ -- yields the relation:
\begin{align}
\frac{|t_i|}{V_i} = \sqrt{\frac{3}{2}\frac{V_2}{V_1}-\frac{3}{4}} \label{e:dpt}
\end{align}
which is shown in the final ground state phase diagram (Fig.~\ref{f:phaseT0}) as dashed line for small $|t_i|/V_i$. Since equation \eqref{e:dpt} does not depend on the sign of $t_i$ it holds also for antiferromagnetic $x$-$y$ interactions and can be compared to the result of a series expansion by Oitmaa {\it et al.}\cite{P:oitmaa96} where for small fluctuations $t_i > 0$  the direct transition between the classical antiferromagnetic states survives as well.

\section{Quantum Monte-Carlo \label{p:QMC}}
\subsection{Algorithm}
For the calculation of the complete phase diagram at finite temperatures and in the ground state we use quantum Monte-Carlo techniques. For negative hopping integrals $t_i$ the Stochastic Series Expansion (SSE)\cite{P:handscomb62, P:sandvik91, P:sandvik92} has no sign problem  and the \emph{directed loop algorithm}\cite{P:sandvik02} yields an adapted update scheme for a large set of parameters in this model. We used an implementation of the ALPS-project\cite{P:alet05, P:ALPS05, P:ALPS07} as basis for our simulations. 

However, the frustration in the model produces a \emph{critical slowing down} in the Monte-Carlo (MC) simulation and the large degeneracy in the vicinity of the critical point $V_2 = V_1 /2$ causes severe thermalization problems in the standard implementation. To overcome these problems we added an \emph{exchange MC} update \cite{P:hukushima96, P:katzgraber06, P:melko07} in temperature space to the ALPS directed-loop application. Starting from the expansion of the partition function in the SSE scheme one can derive an appropriate Metropolis update probability due to detailed balance for this update.\cite{P:melko07}
To guarantee a good thermalization within the exchange MC step it is important to adjust the temperature steps and number of sweeps between the exchanges of configurations (swaps) of neighboring simulations. 

In addition to the exchange MC, we used an annealing procedure for each copy independently during the thermalization process. This kind of algorithm helps to prethermalize the simulations at lower temperatures to ensure a better swap rate for the exchange MC algorithm.

\subsection{Magnetic order}
\begin{figure}
\includegraphics[width=0.48\textwidth]{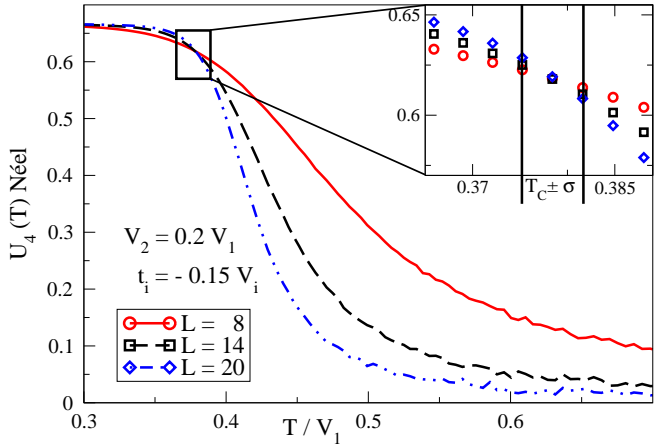}
\caption{\label{f:binder}(Color online) Temperature dependence of the fourth-order cumulants for the N\'eel order parameter for different lattice sizes at $V_2 = 0.2~V_1$ and $t_i = -0.15~V_i$. In the inset the estimate of the error for the transition temperature is given.}
\end{figure}

To determine the regions in the phase diagram where the system is classically ordered, we calculated the structure factors
\begin{align}
S(\vec q) = \frac{1}{N}\sum_{i,j} e^{i \vec q (\vec r_i- \vec r_j)}\langle S_i^z S_j^z \rangle
\end{align} 
for N\'eel order ($\vec q = (\pi,\pi)$) and collinear order ($\vec q = (\pi,0)$ and $(0,\pi)$) for various parameters of the Hamiltonian  \eqref{e:hamil} to check for antiferromagnetic order in the $S^z$ direction. To detect the exact transition temperature into the magnetic phases we took the fourth-order (Binder) cumulants of the order parameter:\cite{P:binder81L, P:binder81Z}
\begin{align}
U_4 = 1 -\frac{\langle m^4 \rangle}{3~\langle m^2 \rangle^2}\,,
\quad m = \sqrt{\frac{S(\vec q)}{N}}
\end{align}
which intersect for different lattice sizes $L$ at the critical temperature. As an example the scenario at $V_2 = 0.2~V_1$ and $t_i = -0.15~V_i$ is shown in Fig.~\ref{f:binder}.
Using this method we constructed the finite-temperature phase-diagrams for different ratios of quantum fluctuations $-t_i/V_i$ for the two antiferromagnetic phases and the high-temperature paramagnetic phase shown in Fig.~\ref{f:phaseT}. The transition temperatures are suppressed to lower values for increasing hopping terms. For $-t_i/V_i \gtrsim 0.1$ and values of $V_2 \approx V_1/2$, i.\,e., in the vicinity of the critical point no magnetic ordering in the $S^z$ direction can be detected. 

We extrapolated the finite-temperature results to $T=0$ to draw the phase boundaries of the magnetic phases in the ground state phase diagram in Fig.~\ref{f:phaseT0}. The shape of the phase diagram is very similar to that calculated by Oitmaa {\it et al.}\cite{P:oitmaa96} for the anisotropic $J_1$-$J_2$-Heisenberg model with antiferromagentic interactions in $S^{x,y,z}$ by means of series expansion in $t_i$ around the Ising limit.

\begin{figure}
\includegraphics[width=0.48\textwidth]{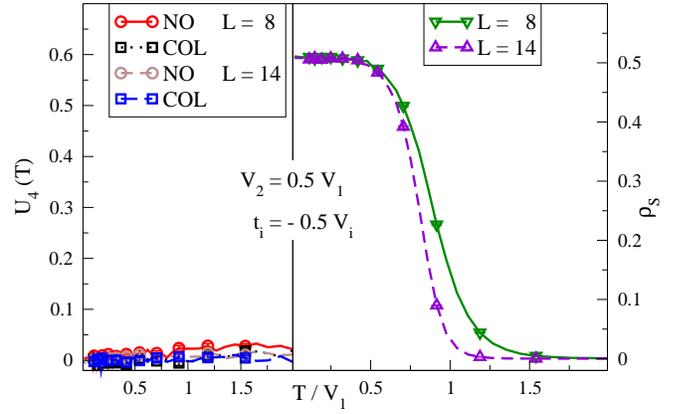}
\caption{\label{f:super}(Color online) Comparison of the superfluid density (spin stiffness) $\rho_S$ (right) and cumulants for magnetic order parameters (left) -- N\'eel order (NO) and collinear order (CO) -- for $-t_i/V_i = 0.5$ and $V_2/V_1 = 0.5$. The clear signature in the superfluid order parameter indicates a finite-temperature phase transition.}
\end{figure}

Furthermore, we performed measurements for the superfluid density (or in spin language the spin stiffness)\cite{P:matsubara56} which is measured via the variance for the net-direction of the off-diagonal operators in the QMC simulations.\cite{P:pollock87} This quantity indicates a correlated movement of the hard-core bosons or a ferromagnetic order in the $x$-$y$ plane respectively. For large $S^x$, $S^y$ interactions $|t_i|$ we find this state to be the ground state. For $-t_i/V_i = 0.5$ and $V_2 = 0.5~V_1$ the temperature dependence of $\rho_S$ is shown in Fig.~\ref{f:super}
for two different lattice sizes $L = 8,~14$.

\begin{figure}
\includegraphics[width=0.48\textwidth]{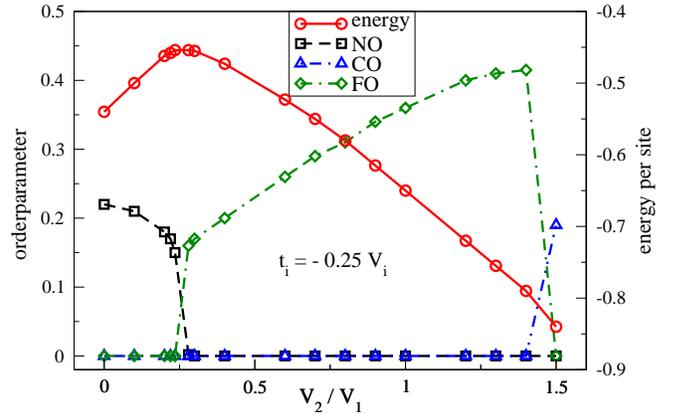}
\caption{\label{f:orderparam05}(Color online)  Ground state values of the order parameters for N\'eel order (NO), collinear order (CO), in-plane ferromagnetic order (FO) and the energy density at $t_i = -0.25~V_i$ and varying frustration $V_2/V_1$. (Symbols are larger than error bars and values do not change for $20 \geq L \geq 16$).}
\end{figure}

\begin{figure}
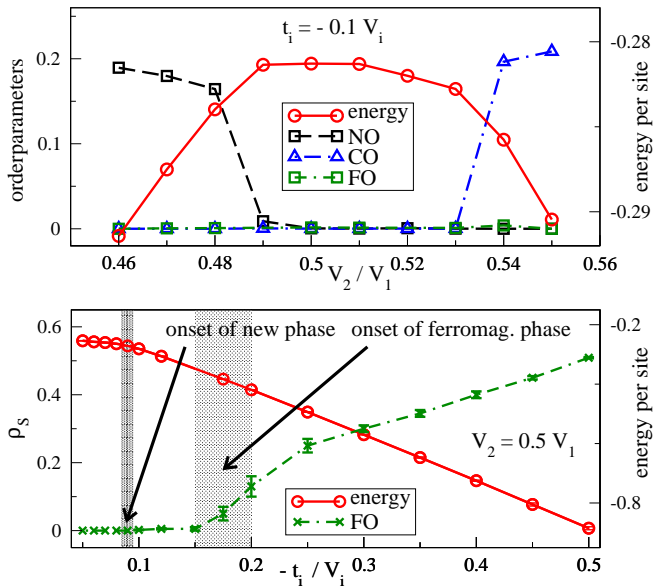

\includegraphics[width=0.48\textwidth]{order02.eps}\\[0.2cm]
\includegraphics[width=0.48\textwidth]{stiff_t.eps}
\caption{\label{f:stiff_t}(Color online) {\bf Top:} Ground-state values of the order parameters for N\'eel order (NO), collinear order (CO), in-plane ferromagnetic order (FO) and the energy density at $t_i = -0.1~V_i$ and varying frustration $V_2/V_1$ (symbols are larger than error-bars). {\bf Bottom:} Evolution of the ferromagnetic order parameter $\rho_S$ at the critical point $V_2 = V_1 / 2$ for increasing ratios $|t_i|/V_i$. For small fluctuations $\rho_S=0$ even after the direct transition between N\'eel and collinear order is suppressed; only for $|t_i|/V_i \gtrsim 0.175$ a finite value of the order parameter is measurable. (Symbols are larger than error bars (if not given) and values do not change for $20 \geq L \geq 16$)}
\end{figure}

In Fig.~\ref{f:orderparam05}
we show the evolution of the order parameters (N\'eel, collinear and in-plane order) for different frustration strengths and fixed quantum fluctuations $-t_i/V_i = 0.25$ at a sufficiently low temperature. The measurements are converged to their ground state values and do not depend on the lattice size any more. The transitions from N\'eel order to in-plane ferromagnetic order and back to antiferromagnetic order (collinear configuration) are clearly visible and we conclude from the sharp features in the order parameters that the transitions are of first order.
However, calculating the observables for smaller values of $|t_i|/V_i$ gives rise to interesting behavior especially for the ferromagnetic order parameter. For the highly frustrated region around $V_2 \approx V_1/2$ and small $|t_i|/V_i$ (with values $0.08 < |t_i|/V_i < 0.175$) the signature of $\rho_S$ depends on the lattice size and goes to zero for low temperatures and large lattices. Thus, we find a small region without any of the conventional order parameters giving a clear sign for an ordering process (Fig.~\ref{f:stiff_t} top).
In addition we show the trend of the ferromagnetic order parameter at $V_2 = V_1 / 2$ in Fig.~\ref{f:stiff_t} (bottom) where the calculations in the intermediate region needed larger lattices and very low temperatures ($20 \geq L \geq 12$, $T \approx 0.01~V_1$) to converge to their ground state value. In smaller lattices a strong tendency to superfluid order was noticed due to the periodic boundary conditions. For increasing lattice size the order parameter began to oscillate before vanishing completely for small temperatures. Since this calculation is very time-consuming we performed it only exemplary for the single value of $V_2= V_1 /2$ and do not give the exact phase transition from the disordered phase into the ferromagnetic phase in Fig.~\ref{f:phaseT0} for all values of $V_2$. However, the smooth behavior of the order parameter and the energy in Fig.~\ref{f:stiff_t} indicates that the transition into the ferromagnetic phase is probably continuous.

\subsection{Quantum correlations}
The lack of the expected conventional order in a finite region of the ground state phase diagram motivated further simulations and calculations of new order parameters. We performed measurements for the whole structure factor and thereby ruled out any magnetic order in the $S^z$-direction. We also checked for unconventional order, i.\,e., quantum-ordered phases like columnar or staggered dimer phases. For this purpose we implemented the measurements for bond-bond correlations which correspond to four-spin correlation functions. In the SSE one can use an improved estimator for correlations between operators that are part of the hamiltonian itself.\cite{P:sandvik92}

\begin{figure}
\includegraphics[width=0.48\textwidth]{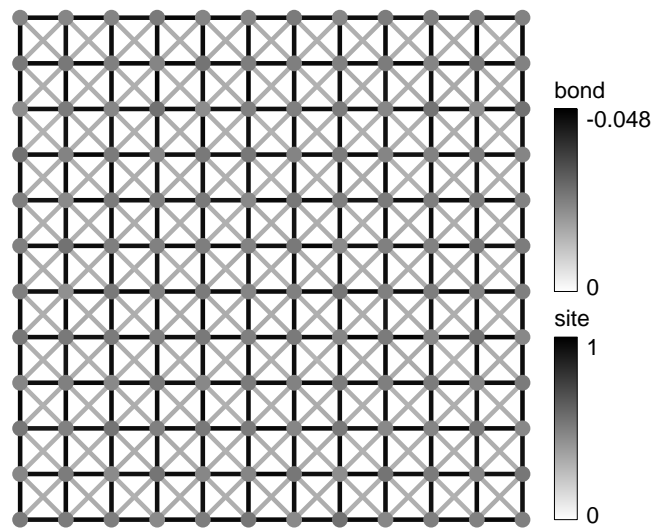}
\caption{\label{f:local}Local measurements of magnetization and kinetic bond energy in a grayscale for a $12 \times 12$ lattice at $V_2 = 0.51~V_1$ and $t_i = -0.1~V_i$ in the ground state ($T=0.01~V_1$). No ordered features are distinguishable.}
\end{figure}

The measurements of local quantities as local magnetization and local kinetic bond energies are shown in Fig.~\ref{f:local}
for parameters which lie in the critical region ($-t_i/V_i = 0.1$ and $V_2 = 0.51~ V_1$) in a grayscale. There is no distinguishable magnetic order and the energies of the bonds are equally distributed. Diagonal bonds have smaller energies due to the fact that $|t_2| < |t_1|$. 

\begin{figure}
\includegraphics[width=0.48\textwidth]{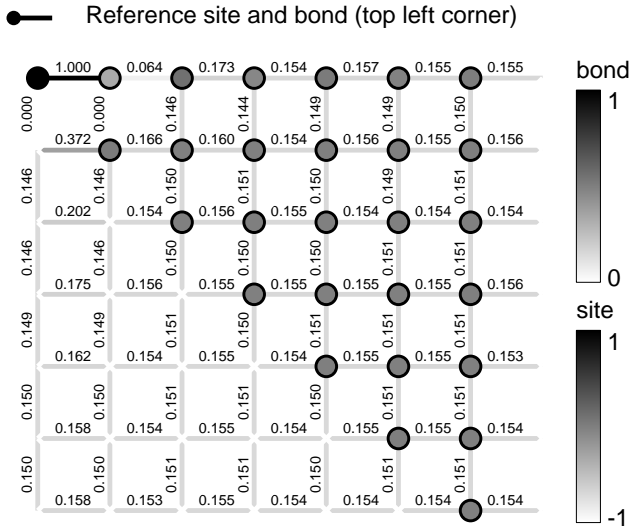}
\caption{\label{f:corrs}We show correlations between spins and dimers living on nearest-neighbor bonds on a periodic $12 \times 12$ lattice, hence up to $6$ neighboring spins and bonds. We chose the top left spin and adjacent right bond as reference points for the measurement. All other sites and bonds of the plot represent the correlation of spins and dimers in the illustrated distance to the reference bond coded in a grayscale. For $V_2 = 0.51~V_1$ and $t_i = -0.1~V_i$ and in the ground state ($T=0.01~V_1$), i.\,e., in the region without conventional order, spin correlations decay exponentially to zero and dimer correlations decay equally to a finite value, hence, we find no long-range order neither in dimer nor in spin correlations.}
\end{figure}

The correlation measurements of spins and bonds are shown in Fig.~\ref{f:corrs}.
As a reference we chose the top left spin with its right adjacent bond. All other bonds and sites represent the strength of the correlation from this special site (bond) to the reference site (bond). For the bond-bond correlations kinetic and potential energy terms are taken into account and they are normalized to the autocorrelation of the reference bond which yields the largest value (black in the figure). There is no long-range order to identify neither in spin-spin correlation which decay exponentially fast nor in the bond-bond correlation which decay all equally to a finite value that is given by the local bond energy. Here the weakest correlations are measured for orthogonal bonds of the same square (top left) where the strongest correlation is given by the parallel bond on the same square. We conclude from these measurements that there does not exist any long-range order in the system for a finite region of parameters.

\section{Exact Diagonalization}

A topologically ordered state could be another possibility for the phase without any signatures for order. To check for this kind of non-local ordering, the calculation of the spectrum is necessary since a degeneracy of the groundstate is  expected for a topologically ordered state.\cite{B:misguich2, P:misguich02} The spectrum is not accessible via QMC simulations and therefore we performed an exact diagonalization for a $6\times 6$ lattice with periodic boundary conditions -- as on a torus -- at $t_i = -0.1~V_i$ and $V_2 = 0.5~V_1$. For the computation of the spectrum we used an existing implementation of an exact diagonalization by J\"org Schulenburg.\footnote{The code for the \emph{spinpack} is available at http://www.ovgu.de/jschulen/spin}

\begin{figure}
\includegraphics[height=0.48\textwidth,angle=270]{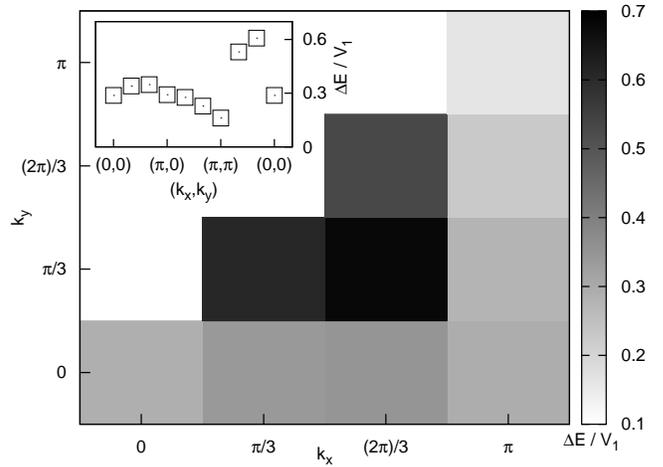}
\caption{\label{f:diag}Exact diagonalization data for a $6\times 6$ lattice
with periodic boundary conditions at $V_2 = 0.5~V_1$ and $t_i = -0.1~V_i$.
Shown are the energy differences $\Delta E = E_k - E_0$ for different
$k$-values (only $k_y \leq k_x$) in a grayscale
($\Delta E_{\min}(\pi,\pi) = 0.1609~V_1$). The spectrum shows no tendency for degeneracy and therefore a topologically ordered state is unlikely. In the inset the same values are shown along the path $(0,0)\rightarrow (\pi,0)\rightarrow (\pi,\pi)\rightarrow (0,0)$.}
\end{figure} 

We calculated the lowest eigenvalues in the different $k$-sectors and show the energy difference to the lowest eigenvalue $E_0(0,0) = -10.2854~V_1$ in Fig.~\ref{f:diag}. For symmetry reasons, it is sufficient to concentrate on the region $0 \le k_x \le \pi$, $0\le k_y \le k_x$ of the Brillouin zone. The main panel of Fig.~\ref{f:diag} shows a grayscale plot of the energy differences in this region, the inset a different representation of essentially the same data. We obtain a minimal energy gap $\Delta E_{\min} = 0.1609~V_1$ for $k=(\pi,\pi)$ above a unique ground state on the $6\times 6$ lattice. This is comparable to the dispersion along the $k_x$-direction where we find a maximum gap $\Delta E(2\pi/3,0) = 0.3482~V_1$. Accordingly, we interpret the large value of $\Delta E_{\min}$ as evidence against a ground-state degeneracy.

The spectrum shown in the inset in Fig.~\ref{f:diag} is also qualitatively different from the spectrum of the Heisenberg model  (cf., e.g., Refs.\ \onlinecite{B:richter04,P:majumdar10}). Indeed, in the latter case one would expect a behavior close to $k_x=0$, $k_y=0$ which is similar to the vicinity of the ordering wave vector (which is $k_x=\pi$, $k_y=\pi$ for the N\'eel state). The excitation spectrum shown in Fig.~\ref{f:diag} is therefore consistent with the absence of classical order as demonstrated by the QMC simulations. 

While it would be desirable to perform a finite-size analysis, we have selected the $6 \times 6$ lattice as the only accessible lattice which has the symmetries of the infinite system and is compatible with the expected ordered states (the $4 \times 4$ `square lattice' is not necessarily representative for two dimensions since it can also be interpreted as a four-dimensional torus). Still, the gap $\Delta E_{\min}$ is comparable to the dispersion of the excited states even on the $6 \times 6$ lattice which suggests that the gap will also stay finite in the thermodynamic limit. According to this result no groundstate degeneracy and for this reason no topological order is present.

\section{Discussion}
We simulated a two-dimensional lattice model for hard-core bosons with competing interactions which maps onto an anisotropic frustrated quantum spin model. We used improved QMC methods to calculate classical magnetic order parameters for a large parameter space of the model. We found two antiferromagnetic phases as expected from the classical limit (the frustrated Ising model) and a ferromagnetic configuration which corresponds to a superfluid phase in the bosonic language. Most importantly, we detected a finite region without any magnetic order. Careful calculations of higher correlations as four-spin correlations to check for dimer ordered phases which were proposed by Balents {\it et al.} for similar models \cite{P:bartosch05} did not give any signal of long-ranged order. Further exact diagonalization calculations showed that the system is not topologically ordered, either. In conclusion, there seems to exist a finite region in parameter space without any long-range order and with an exponential decay in the spin-spin correlations which has not been observed in earlier works on similar models.\cite{P:batrouni00, P:batrouni01, P:melko08, P:chen08} Since we did not find any dimer-ordered phases we assume that their appearence strongly depends on the kinetic energy encoded in the model. An introduction of different exchange terms as, e.\,g., a ring exchange could enhance the probability of finding valence bond crystals.\cite{P:bartosch05,P:sandvik02b}

\begin{acknowledgments}
We would like to thank the Deutsche Forschungsgemeinschaft for financial support via the collaborative research center SFB 602 (TP 18) and a Heisenberg fellowship (grant \# 2325/4-2, A. Honecker). Furthermore we would like to mention that most of the QMC simulations were computed on the parallel clusters of the North-German Supercomputing Alliance (HLRN) and would like to thank them for technical support. In addition we would like to thank Dr. Jörg Schulenburg for providing us with his diagonalization code.
\end{acknowledgments}

\bibliography{../Literatur}

\begin{thebibliography}{49}%
\makeatletter
\providecommand \@ifxundefined [1]{%
 \@ifx{#1\undefined}
}%
\providecommand \@ifnum [1]{%
 \ifnum #1\expandafter \@firstoftwo
 \else \expandafter \@secondoftwo
 \fi
}%
\providecommand \@ifx [1]{%
 \ifx #1\expandafter \@firstoftwo
 \else \expandafter \@secondoftwo
 \fi
}%
\providecommand \natexlab [1]{#1}%
\providecommand \enquote  [1]{``#1''}%
\providecommand \bibnamefont  [1]{#1}%
\providecommand \bibfnamefont [1]{#1}%
\providecommand \citenamefont [1]{#1}%
\providecommand \href@noop [0]{\@secondoftwo}%
\providecommand \href [0]{\begingroup \@sanitize@url \@href}%
\providecommand \@href[1]{\@@startlink{#1}\@@href}%
\providecommand \@@href[1]{\endgroup#1\@@endlink}%
\providecommand \@sanitize@url [0]{\catcode `\\12\catcode `\$12\catcode
  `\&12\catcode `\#12\catcode `\^12\catcode `\_12\catcode `\%12\relax}%
\providecommand \@@startlink[1]{}%
\providecommand \@@endlink[0]{}%
\providecommand \url  [0]{\begingroup\@sanitize@url \@url }%
\providecommand \@url [1]{\endgroup\@href {#1}{\urlprefix }}%
\providecommand \urlprefix  [0]{URL }%
\providecommand \Eprint [0]{\href }%
\providecommand \doibase [0]{http://dx.doi.org/}%
\providecommand \selectlanguage [0]{\@gobble}%
\providecommand \bibinfo  [0]{\@secondoftwo}%
\providecommand \bibfield  [0]{\@secondoftwo}%
\providecommand \translation [1]{[#1]}%
\providecommand \BibitemOpen [0]{}%
\providecommand \bibitemStop [0]{}%
\providecommand \bibitemNoStop [0]{.\EOS\space}%
\providecommand \EOS [0]{\spacefactor3000\relax}%
\providecommand \BibitemShut  [1]{\csname bibitem#1\endcsname}%
\let\auto@bib@innerbib\@empty
%</preamble>
\bibitem [{\citenamefont {Misguich}\ and\ \citenamefont
  {Lhuillier}(2005)}]{B:misguich2}%
  \BibitemOpen
  \bibfield  {author} {\bibinfo {author} {\bibfnamefont {G.}~\bibnamefont
  {Misguich}}\ and\ \bibinfo {author} {\bibfnamefont {C.}~\bibnamefont
  {Lhuillier}},\ }\enquote {\bibinfo {title} {Two-dimensional quantum
  antiferromagnets},}\ in\ \href@noop {} {\emph {\bibinfo {booktitle}
  {Frustrated spin systems}}},\ \bibinfo {editor} {edited by\ \bibinfo {editor}
  {\bibfnamefont {H.~T.}\ \bibnamefont {Diep}}}\ (\bibinfo  {publisher} {World
  Scientific},\ \bibinfo {year} {2005})\ Chap.~\bibinfo {chapter}
  {5.}\BibitemShut {Stop}%
\bibitem [{\citenamefont {Lacroix}\ \emph {et~al.}(2011)\citenamefont
  {Lacroix}, \citenamefont {Mendels},\ and\ \citenamefont {Mila}}]{B:mila11}%
  \BibitemOpen
  \bibinfo {editor} {\bibfnamefont {C.}~\bibnamefont {Lacroix}}, \bibinfo
  {editor} {\bibfnamefont {P.}~\bibnamefont {Mendels}}, \ and\ \bibinfo
  {editor} {\bibfnamefont {F.}~\bibnamefont {Mila}},\ eds.,\ \href
  {http://www.springer.com/materials/book/978-3-642-10588-3} {\emph {\bibinfo
  {title} {Introduction to Frustrated Magnetism}}},\ \bibinfo {edition} {1st}\
  ed.,\ \bibinfo {series} {Springer Series in Solid-State Sciences}, Vol.\
  \bibinfo {volume} {164}\ (\bibinfo  {publisher} {Springer},\ \bibinfo {year}
  {2011})\BibitemShut {NoStop}%
\bibitem [{\citenamefont {Capriotti}\ \emph {et~al.}(2003)\citenamefont
  {Capriotti}, \citenamefont {Becca}, \citenamefont {Parola},\ and\
  \citenamefont {Sorella}}]{P:capriotti03}%
  \BibitemOpen
  \bibfield  {author} {\bibinfo {author} {\bibfnamefont {L.}~\bibnamefont
  {Capriotti}}, \bibinfo {author} {\bibfnamefont {F.}~\bibnamefont {Becca}},
  \bibinfo {author} {\bibfnamefont {A.}~\bibnamefont {Parola}}, \ and\ \bibinfo
  {author} {\bibfnamefont {S.}~\bibnamefont {Sorella}},\ }\href {\doibase
  10.1103/PhysRevB.67.212402} {\bibfield  {journal} {\bibinfo  {journal} {Phys.
  Rev. B}\ }\textbf {\bibinfo {volume} {67}},\ \bibinfo {pages} {212402}
  (\bibinfo {year} {2003})}\BibitemShut {NoStop}%
\bibitem [{\citenamefont {Richter}\ \emph {et~al.}(2004)\citenamefont
  {Richter}, \citenamefont {Schulenburg},\ and\ \citenamefont
  {Honecker}}]{B:richter04}%
  \BibitemOpen
  \bibfield  {author} {\bibinfo {author} {\bibfnamefont {J.}~\bibnamefont
  {Richter}}, \bibinfo {author} {\bibfnamefont {J.}~\bibnamefont
  {Schulenburg}}, \ and\ \bibinfo {author} {\bibfnamefont {A.}~\bibnamefont
  {Honecker}},\ }in\ \href@noop {} {\emph {\bibinfo {booktitle} {Quantum
  magnetism}}},\ \bibinfo {series} {Lecture Notes in Physics}, Vol.\ \bibinfo
  {volume} {645},\ \bibinfo {editor} {edited by\ \bibinfo {editor}
  {\bibfnamefont {U.}~\bibnamefont {Schollw\"ock}}, \bibinfo {editor}
  {\bibfnamefont {J.}~\bibnamefont {Richter}}, \bibinfo {editor} {\bibfnamefont
  {D.~J.~J.}\ \bibnamefont {Farnell}}, \ and\ \bibinfo {editor} {\bibfnamefont
  {R.~F.}\ \bibnamefont {Bishop}}}\ (\bibinfo  {publisher} {Springer},\
  \bibinfo {year} {2004})\ Chap.~\bibinfo {chapter} {2}, p.~\bibinfo {pages}
  {85}\BibitemShut {NoStop}%
\bibitem [{\citenamefont {Mambrini}\ \emph {et~al.}(2006)\citenamefont
  {Mambrini}, \citenamefont {L\"auchli}, \citenamefont {Poilblanc},\ and\
  \citenamefont {Mila}}]{P:mambrini06}%
  \BibitemOpen
  \bibfield  {author} {\bibinfo {author} {\bibfnamefont {M.}~\bibnamefont
  {Mambrini}}, \bibinfo {author} {\bibfnamefont {A.}~\bibnamefont {L\"auchli}},
  \bibinfo {author} {\bibfnamefont {D.}~\bibnamefont {Poilblanc}}, \ and\
  \bibinfo {author} {\bibfnamefont {F.}~\bibnamefont {Mila}},\ }\href {\doibase
  10.1103/PhysRevB.74.144422} {\bibfield  {journal} {\bibinfo  {journal} {Phys.
  Rev. B}\ }\textbf {\bibinfo {volume} {74}},\ \bibinfo {pages} {144422}
  (\bibinfo {year} {2006})}\BibitemShut {NoStop}%
\bibitem [{\citenamefont {Bishop}\ \emph {et~al.}(2008)\citenamefont {Bishop},
  \citenamefont {Li}, \citenamefont {Darradi},\ and\ \citenamefont
  {Richter}}]{P:darradi08}%
  \BibitemOpen
  \bibfield  {author} {\bibinfo {author} {\bibfnamefont {R.~F.}\ \bibnamefont
  {Bishop}}, \bibinfo {author} {\bibfnamefont {P.}~\bibnamefont {Li}}, \bibinfo
  {author} {\bibfnamefont {R.}~\bibnamefont {Darradi}}, \ and\ \bibinfo
  {author} {\bibfnamefont {J.}~\bibnamefont {Richter}},\ }\href {\doibase
  10.1088/0953-8984/20/25/255251} {\bibfield  {journal} {\bibinfo  {journal}
  {J. Phys.: Cond. Matt.}\ }\textbf {\bibinfo {volume} {20}},\ \bibinfo {pages}
  {255251} (\bibinfo {year} {2008})}\BibitemShut {NoStop}%
\bibitem [{\citenamefont {Ralko}\ \emph {et~al.}(2009)\citenamefont {Ralko},
  \citenamefont {Mambrini},\ and\ \citenamefont {Poilblanc}}]{P:ralko09}%
  \BibitemOpen
  \bibfield  {author} {\bibinfo {author} {\bibfnamefont {A.}~\bibnamefont
  {Ralko}}, \bibinfo {author} {\bibfnamefont {M.}~\bibnamefont {Mambrini}}, \
  and\ \bibinfo {author} {\bibfnamefont {D.}~\bibnamefont {Poilblanc}},\ }\href
  {\doibase 10.1103/PhysRevB.80.184427} {\bibfield  {journal} {\bibinfo
  {journal} {Phys. Rev. B}\ }\textbf {\bibinfo {volume} {80}},\ \bibinfo
  {pages} {184427} (\bibinfo {year} {2009})}\BibitemShut {NoStop}%
\bibitem [{\citenamefont {Richter}\ \emph {et~al.}(2010)\citenamefont
  {Richter}, \citenamefont {Darradi}, \citenamefont {Schulenburg},
  \citenamefont {Farnell},\ and\ \citenamefont {Rosner}}]{P:richter10}%
  \BibitemOpen
  \bibfield  {author} {\bibinfo {author} {\bibfnamefont {J.}~\bibnamefont
  {Richter}}, \bibinfo {author} {\bibfnamefont {R.}~\bibnamefont {Darradi}},
  \bibinfo {author} {\bibfnamefont {J.}~\bibnamefont {Schulenburg}}, \bibinfo
  {author} {\bibfnamefont {D.~J.~J.}\ \bibnamefont {Farnell}}, \ and\ \bibinfo
  {author} {\bibfnamefont {H.}~\bibnamefont {Rosner}},\ }\href {\doibase
  10.1103/PhysRevB.81.174429} {\bibfield  {journal} {\bibinfo  {journal} {Phys.
  Rev. B}\ }\textbf {\bibinfo {volume} {81}},\ \bibinfo {pages} {174429}
  (\bibinfo {year} {2010})}\BibitemShut {NoStop}%
\bibitem [{\citenamefont {Richter}\ and\ \citenamefont
  {Schulenburg}(2010)}]{P:richter10b}%
  \BibitemOpen
  \bibfield  {author} {\bibinfo {author} {\bibfnamefont {J.}~\bibnamefont
  {Richter}}\ and\ \bibinfo {author} {\bibfnamefont {J.}~\bibnamefont
  {Schulenburg}},\ }\href {\doibase 10.1140/epjb/e2009-00400-4} {\bibfield
  {journal} {\bibinfo  {journal} {Eur. Phys. J. B}\ }\textbf {\bibinfo {volume}
  {73}},\ \bibinfo {pages} {117} (\bibinfo {year} {2010})}\BibitemShut
  {NoStop}%
\bibitem [{\citenamefont {Batrouni}\ and\ \citenamefont
  {Scalettar}(2000)}]{P:batrouni00}%
  \BibitemOpen
  \bibfield  {author} {\bibinfo {author} {\bibfnamefont {G.~G.}\ \bibnamefont
  {Batrouni}}\ and\ \bibinfo {author} {\bibfnamefont {R.~T.}\ \bibnamefont
  {Scalettar}},\ }\href {\doibase 10.1103/PhysRevLett.84.1599} {\bibfield
  {journal} {\bibinfo  {journal} {Phys. Rev. Lett.}\ }\textbf {\bibinfo
  {volume} {84}},\ \bibinfo {pages} {1599} (\bibinfo {year}
  {2000})}\BibitemShut {NoStop}%
\bibitem [{\citenamefont {H\'ebert}\ \emph {et~al.}(2001)\citenamefont
  {H\'ebert}, \citenamefont {Batrouni}, \citenamefont {Scalettar},
  \citenamefont {Schmid}, \citenamefont {Troyer},\ and\ \citenamefont
  {Dorneich}}]{P:batrouni01}%
  \BibitemOpen
  \bibfield  {author} {\bibinfo {author} {\bibfnamefont {F.}~\bibnamefont
  {H\'ebert}}, \bibinfo {author} {\bibfnamefont {G.~G.}\ \bibnamefont
  {Batrouni}}, \bibinfo {author} {\bibfnamefont {R.~T.}\ \bibnamefont
  {Scalettar}}, \bibinfo {author} {\bibfnamefont {G.}~\bibnamefont {Schmid}},
  \bibinfo {author} {\bibfnamefont {M.}~\bibnamefont {Troyer}}, \ and\ \bibinfo
  {author} {\bibfnamefont {A.}~\bibnamefont {Dorneich}},\ }\href {\doibase
  10.1103/PhysRevB.65.014513} {\bibfield  {journal} {\bibinfo  {journal} {Phys.
  Rev. B}\ }\textbf {\bibinfo {volume} {65}},\ \bibinfo {pages} {014513}
  (\bibinfo {year} {2001})}\BibitemShut {NoStop}%
\bibitem [{\citenamefont {Ng}\ and\ \citenamefont {Chen}(2008)}]{P:chen08}%
  \BibitemOpen
  \bibfield  {author} {\bibinfo {author} {\bibfnamefont {K.~K.}\ \bibnamefont
  {Ng}}\ and\ \bibinfo {author} {\bibfnamefont {Y.~C.}\ \bibnamefont {Chen}},\
  }\href {\doibase 10.1103/PhysRevB.77.052506} {\bibfield  {journal} {\bibinfo
  {journal} {Phys. Rev. B}\ }\textbf {\bibinfo {volume} {77}},\ \bibinfo {eid}
  {052506} (\bibinfo {year} {2008})}\BibitemShut {NoStop}%
\bibitem [{\citenamefont {Chen}\ \emph {et~al.}(2008)\citenamefont {Chen},
  \citenamefont {Melko}, \citenamefont {Wessel},\ and\ \citenamefont
  {Kao}}]{P:melko08}%
  \BibitemOpen
  \bibfield  {author} {\bibinfo {author} {\bibfnamefont {Y.~C.}\ \bibnamefont
  {Chen}}, \bibinfo {author} {\bibfnamefont {R.~G.}\ \bibnamefont {Melko}},
  \bibinfo {author} {\bibfnamefont {S.}~\bibnamefont {Wessel}}, \ and\ \bibinfo
  {author} {\bibfnamefont {Y.~J.}\ \bibnamefont {Kao}},\ }\href {\doibase
  10.1103/PhysRevB.77.014524} {\bibfield  {journal} {\bibinfo  {journal} {Phys.
  Rev. B}\ }\textbf {\bibinfo {volume} {77}},\ \bibinfo {eid} {014524}
  (\bibinfo {year} {2008})}\BibitemShut {NoStop}%
\bibitem [{\citenamefont {Landau}(1980)}]{P:landau80}%
  \BibitemOpen
  \bibfield  {author} {\bibinfo {author} {\bibfnamefont {D.~P.}\ \bibnamefont
  {Landau}},\ }\href {\doibase 10.1103/PhysRevB.21.1285} {\bibfield  {journal}
  {\bibinfo  {journal} {Phys. Rev. B}\ }\textbf {\bibinfo {volume} {21}},\
  \bibinfo {pages} {1285} (\bibinfo {year} {1980})}\BibitemShut {NoStop}%
\bibitem [{\citenamefont {Binder}\ and\ \citenamefont
  {Landau}(1980)}]{P:lanbin80}%
  \BibitemOpen
  \bibfield  {author} {\bibinfo {author} {\bibfnamefont {K.}~\bibnamefont
  {Binder}}\ and\ \bibinfo {author} {\bibfnamefont {D.~P.}\ \bibnamefont
  {Landau}},\ }\href {\doibase 10.1103/PhysRevB.21.1941} {\bibfield  {journal}
  {\bibinfo  {journal} {Phys. Rev. B}\ }\textbf {\bibinfo {volume} {21}},\
  \bibinfo {pages} {1941} (\bibinfo {year} {1980})}\BibitemShut {NoStop}%
\bibitem [{\citenamefont {Landau}\ and\ \citenamefont
  {Binder}(1985)}]{P:lanbin85}%
  \BibitemOpen
  \bibfield  {author} {\bibinfo {author} {\bibfnamefont {D.~P.}\ \bibnamefont
  {Landau}}\ and\ \bibinfo {author} {\bibfnamefont {K.}~\bibnamefont
  {Binder}},\ }\href {\doibase 10.1103/PhysRevB.31.5946} {\bibfield  {journal}
  {\bibinfo  {journal} {Phys. Rev. B}\ }\textbf {\bibinfo {volume} {31}},\
  \bibinfo {pages} {5946} (\bibinfo {year} {1985})}\BibitemShut {NoStop}%
\bibitem [{\citenamefont {Mor\'an-L\'opez}\ \emph {et~al.}(1993)\citenamefont
  {Mor\'an-L\'opez}, \citenamefont {Aguilera-Granja},\ and\ \citenamefont
  {Sanchez}}]{P:lopez93}%
  \BibitemOpen
  \bibfield  {author} {\bibinfo {author} {\bibfnamefont {J.~L.}\ \bibnamefont
  {Mor\'an-L\'opez}}, \bibinfo {author} {\bibfnamefont {F.}~\bibnamefont
  {Aguilera-Granja}}, \ and\ \bibinfo {author} {\bibfnamefont {J.~M.}\
  \bibnamefont {Sanchez}},\ }\href {\doibase 10.1103/PhysRevB.48.3519}
  {\bibfield  {journal} {\bibinfo  {journal} {Phys. Rev. B}\ }\textbf {\bibinfo
  {volume} {48}},\ \bibinfo {pages} {3519} (\bibinfo {year}
  {1993})}\BibitemShut {NoStop}%
\bibitem [{\citenamefont {Mor\'an-L\'opez}\ \emph {et~al.}(1994)\citenamefont
  {Mor\'an-L\'opez}, \citenamefont {Aguilera-Granja},\ and\ \citenamefont
  {Sanchez}}]{P:lopez94}%
  \BibitemOpen
  \bibfield  {author} {\bibinfo {author} {\bibfnamefont {J.~L.}\ \bibnamefont
  {Mor\'an-L\'opez}}, \bibinfo {author} {\bibfnamefont {F.}~\bibnamefont
  {Aguilera-Granja}}, \ and\ \bibinfo {author} {\bibfnamefont {J.~M.}\
  \bibnamefont {Sanchez}},\ }\href {\doibase 10.1088/0953-8984/6/45/025}
  {\bibfield  {journal} {\bibinfo  {journal} {J. Phys.: Cond. Matt.}\ }\textbf
  {\bibinfo {volume} {6}},\ \bibinfo {pages} {9759} (\bibinfo {year}
  {1994})}\BibitemShut {NoStop}%
\bibitem [{\citenamefont {Malakis}\ \emph {et~al.}(2006)\citenamefont
  {Malakis}, \citenamefont {Kalozoumis},\ and\ \citenamefont
  {Tyraskis}}]{P:malakis06}%
  \BibitemOpen
  \bibfield  {author} {\bibinfo {author} {\bibfnamefont {A.}~\bibnamefont
  {Malakis}}, \bibinfo {author} {\bibfnamefont {P.}~\bibnamefont {Kalozoumis}},
  \ and\ \bibinfo {author} {\bibfnamefont {N.}~\bibnamefont {Tyraskis}},\
  }\href {\doibase 10.1140/epjb/e2006-00032-2} {\bibfield  {journal} {\bibinfo
  {journal} {Eur. Phys. J. B}\ }\textbf {\bibinfo {volume} {50}},\ \bibinfo
  {pages} {63} (\bibinfo {year} {2006})}\BibitemShut {NoStop}%
\bibitem [{\citenamefont {Kalz}\ \emph {et~al.}(2008)\citenamefont {Kalz},
  \citenamefont {Honecker}, \citenamefont {Fuchs},\ and\ \citenamefont
  {Pruschke}}]{P:kalz08}%
  \BibitemOpen
  \bibfield  {author} {\bibinfo {author} {\bibfnamefont {A.}~\bibnamefont
  {Kalz}}, \bibinfo {author} {\bibfnamefont {A.}~\bibnamefont {Honecker}},
  \bibinfo {author} {\bibfnamefont {S.}~\bibnamefont {Fuchs}}, \ and\ \bibinfo
  {author} {\bibfnamefont {T.}~\bibnamefont {Pruschke}},\ }\href {\doibase
  10.1140/epjb/e2008-00359-6} {\bibfield  {journal} {\bibinfo  {journal} {Eur.
  Phys. J. B}\ }\textbf {\bibinfo {volume} {65}},\ \bibinfo {pages} {533}
  (\bibinfo {year} {2008})}\BibitemShut {NoStop}%
\bibitem [{\citenamefont {Kalz}\ \emph {et~al.}(2009)\citenamefont {Kalz},
  \citenamefont {Honecker}, \citenamefont {Fuchs},\ and\ \citenamefont
  {Pruschke}}]{P:kalz09}%
  \BibitemOpen
  \bibfield  {author} {\bibinfo {author} {\bibfnamefont {A.}~\bibnamefont
  {Kalz}}, \bibinfo {author} {\bibfnamefont {A.}~\bibnamefont {Honecker}},
  \bibinfo {author} {\bibfnamefont {S.}~\bibnamefont {Fuchs}}, \ and\ \bibinfo
  {author} {\bibfnamefont {T.}~\bibnamefont {Pruschke}},\ }\href {\doibase
  10.1088/1742-6596/145/1/012051} {\bibfield  {journal} {\bibinfo  {journal}
  {J. Phys.: Conf. Ser.}\ }\textbf {\bibinfo {volume} {145}},\ \bibinfo {pages}
  {012051} (\bibinfo {year} {2009})}\BibitemShut {NoStop}%
\bibitem [{\citenamefont {Bloch}(1932)}]{P:bloch31}%
  \BibitemOpen
  \bibfield  {author} {\bibinfo {author} {\bibfnamefont {F.}~\bibnamefont
  {Bloch}},\ }\href {\doibase 10.1007/BF01337791} {\bibfield  {journal}
  {\bibinfo  {journal} {Z. Phys.}\ }\textbf {\bibinfo {volume} {74}},\ \bibinfo
  {pages} {295} (\bibinfo {year} {1932})}\BibitemShut {NoStop}%
\bibitem [{\citenamefont {H\"ohler}(1950)}]{P:hoehler50}%
  \BibitemOpen
  \bibfield  {author} {\bibinfo {author} {\bibfnamefont {G.}~\bibnamefont
  {H\"ohler}},\ }\href {\doibase 10.1002/andp.19504420110} {\bibfield
  {journal} {\bibinfo  {journal} {Ann. Phys. (Leipzig)}\ }\textbf {\bibinfo
  {volume} {442}},\ \bibinfo {pages} {93} (\bibinfo {year} {1950})}\BibitemShut
  {NoStop}%
\bibitem [{\citenamefont {Matsubara}\ and\ \citenamefont
  {Matsuda}(1956)}]{P:matsubara56}%
  \BibitemOpen
  \bibfield  {author} {\bibinfo {author} {\bibfnamefont {T.}~\bibnamefont
  {Matsubara}}\ and\ \bibinfo {author} {\bibfnamefont {H.}~\bibnamefont
  {Matsuda}},\ }\href {\doibase 10.1143/PTP.16.569} {\bibfield  {journal}
  {\bibinfo  {journal} {Prog. Theor. Phys.}\ }\textbf {\bibinfo {volume}
  {16}},\ \bibinfo {pages} {569} (\bibinfo {year} {1956})}\BibitemShut
  {NoStop}%
\bibitem [{\citenamefont {Liu}\ and\ \citenamefont {Fisher}(1973)}]{P:liu72}%
  \BibitemOpen
  \bibfield  {author} {\bibinfo {author} {\bibfnamefont {K.-S.}\ \bibnamefont
  {Liu}}\ and\ \bibinfo {author} {\bibfnamefont {M.~E.}\ \bibnamefont
  {Fisher}},\ }\href {\doibase 10.1007/BF00655458} {\bibfield  {journal}
  {\bibinfo  {journal} {J. Low Temp. Phys.}\ }\textbf {\bibinfo {volume}
  {10}},\ \bibinfo {pages} {655} (\bibinfo {year} {1973})}\BibitemShut
  {NoStop}%
\bibitem [{\citenamefont {Landau}\ and\ \citenamefont
  {Binder}(2005)}]{B:lanbin00}%
  \BibitemOpen
  \bibfield  {author} {\bibinfo {author} {\bibfnamefont {D.~P.}\ \bibnamefont
  {Landau}}\ and\ \bibinfo {author} {\bibfnamefont {K.}~\bibnamefont
  {Binder}},\ }\href@noop {} {\emph {\bibinfo {title} {Monte Carlo Simulations
  in Statistical Physics}}},\ \bibinfo {edition} {2nd}\ ed.\ (\bibinfo
  {publisher} {Cambridge University Press},\ \bibinfo {year}
  {2005})\BibitemShut {NoStop}%
\bibitem [{\citenamefont {Balents}\ \emph {et~al.}(2005)\citenamefont
  {Balents}, \citenamefont {Bartosch}, \citenamefont {Burkov}, \citenamefont
  {Sachdev},\ and\ \citenamefont {Sengupta}}]{P:bartosch05}%
  \BibitemOpen
  \bibfield  {author} {\bibinfo {author} {\bibfnamefont {L.}~\bibnamefont
  {Balents}}, \bibinfo {author} {\bibfnamefont {L.}~\bibnamefont {Bartosch}},
  \bibinfo {author} {\bibfnamefont {A.}~\bibnamefont {Burkov}}, \bibinfo
  {author} {\bibfnamefont {S.}~\bibnamefont {Sachdev}}, \ and\ \bibinfo
  {author} {\bibfnamefont {K.}~\bibnamefont {Sengupta}},\ }\href {\doibase
  10.1103/PhysRevB.71.144508} {\bibfield  {journal} {\bibinfo  {journal} {Phys.
  Rev. B}\ }\textbf {\bibinfo {volume} {71}},\ \bibinfo {eid} {144508}
  (\bibinfo {year} {2005})}\BibitemShut {NoStop}%
\bibitem [{\citenamefont {Wessel}\ and\ \citenamefont
  {Troyer}(2005)}]{P:wessel05}%
  \BibitemOpen
  \bibfield  {author} {\bibinfo {author} {\bibfnamefont {S.}~\bibnamefont
  {Wessel}}\ and\ \bibinfo {author} {\bibfnamefont {M.}~\bibnamefont
  {Troyer}},\ }\href {\doibase 10.1103/PhysRevLett.95.127205} {\bibfield
  {journal} {\bibinfo  {journal} {Phys. Rev. Lett.}\ }\textbf {\bibinfo
  {volume} {95}},\ \bibinfo {eid} {127205} (\bibinfo {year}
  {2005})}\BibitemShut {NoStop}%
\bibitem [{\citenamefont {Heidarian}\ and\ \citenamefont
  {Damle}(2005)}]{P:damle05}%
  \BibitemOpen
  \bibfield  {author} {\bibinfo {author} {\bibfnamefont {D.}~\bibnamefont
  {Heidarian}}\ and\ \bibinfo {author} {\bibfnamefont {K.}~\bibnamefont
  {Damle}},\ }\href {\doibase 10.1103/PhysRevLett.95.127206} {\bibfield
  {journal} {\bibinfo  {journal} {Phys. Rev. Lett.}\ }\textbf {\bibinfo
  {volume} {95}},\ \bibinfo {pages} {127206} (\bibinfo {year}
  {2005})}\BibitemShut {NoStop}%
\bibitem [{\citenamefont {Melko}\ \emph {et~al.}(2005)\citenamefont {Melko},
  \citenamefont {Paramekanti}, \citenamefont {Burkov}, \citenamefont
  {Vishwanath}, \citenamefont {Sheng},\ and\ \citenamefont
  {Balents}}]{P:melko05}%
  \BibitemOpen
  \bibfield  {author} {\bibinfo {author} {\bibfnamefont {R.~G.}\ \bibnamefont
  {Melko}}, \bibinfo {author} {\bibfnamefont {A.}~\bibnamefont {Paramekanti}},
  \bibinfo {author} {\bibfnamefont {A.~A.}\ \bibnamefont {Burkov}}, \bibinfo
  {author} {\bibfnamefont {A.}~\bibnamefont {Vishwanath}}, \bibinfo {author}
  {\bibfnamefont {D.~N.}\ \bibnamefont {Sheng}}, \ and\ \bibinfo {author}
  {\bibfnamefont {L.}~\bibnamefont {Balents}},\ }\href {\doibase
  10.1103/PhysRevLett.95.127207} {\bibfield  {journal} {\bibinfo  {journal}
  {Phys. Rev. Lett.}\ }\textbf {\bibinfo {volume} {95}},\ \bibinfo {eid}
  {127207} (\bibinfo {year} {2005})}\BibitemShut {NoStop}%
\bibitem [{\citenamefont {Suzuki}\ and\ \citenamefont
  {Okamoto}(1983)}]{P:suzuki83}%
  \BibitemOpen
  \bibfield  {author} {\bibinfo {author} {\bibfnamefont {K.}~\bibnamefont
  {Suzuki}}\ and\ \bibinfo {author} {\bibfnamefont {R.}~\bibnamefont
  {Okamoto}},\ }\href {\doibase 10.1143/PTP.70.439} {\bibfield  {journal}
  {\bibinfo  {journal} {Prog. Theor. Phys.}\ }\textbf {\bibinfo {volume}
  {70}},\ \bibinfo {pages} {439} (\bibinfo {year} {1983})}\BibitemShut
  {NoStop}%
\bibitem [{\citenamefont {Oitmaa}\ and\ \citenamefont
  {Weihong}(1996)}]{P:oitmaa96}%
  \BibitemOpen
  \bibfield  {author} {\bibinfo {author} {\bibfnamefont {J.}~\bibnamefont
  {Oitmaa}}\ and\ \bibinfo {author} {\bibfnamefont {Z.}~\bibnamefont
  {Weihong}},\ }\href {\doibase 10.1103/PhysRevB.54.3022} {\bibfield  {journal}
  {\bibinfo  {journal} {Phys. Rev. B}\ }\textbf {\bibinfo {volume} {54}},\
  \bibinfo {pages} {3022} (\bibinfo {year} {1996})}\BibitemShut {NoStop}%
\bibitem [{\citenamefont {Handscomb}(1962)}]{P:handscomb62}%
  \BibitemOpen
  \bibfield  {author} {\bibinfo {author} {\bibfnamefont {D.}~\bibnamefont
  {Handscomb}},\ }\href {\doibase 10.1017/S0305004100040639} {\bibfield
  {journal} {\bibinfo  {journal} {Proc. Cambr. Philos. Soc.}\ }\textbf
  {\bibinfo {volume} {58}},\ \bibinfo {pages} {594} (\bibinfo {year}
  {1962})}\BibitemShut {NoStop}%
\bibitem [{\citenamefont {Sandvik}\ and\ \citenamefont
  {Kurkij\"arvi}(1991)}]{P:sandvik91}%
  \BibitemOpen
  \bibfield  {author} {\bibinfo {author} {\bibfnamefont {A.~W.}\ \bibnamefont
  {Sandvik}}\ and\ \bibinfo {author} {\bibfnamefont {J.}~\bibnamefont
  {Kurkij\"arvi}},\ }\href {\doibase 10.1103/PhysRevB.43.5950} {\bibfield
  {journal} {\bibinfo  {journal} {Phys. Rev. B}\ }\textbf {\bibinfo {volume}
  {43}},\ \bibinfo {pages} {5950} (\bibinfo {year} {1991})}\BibitemShut
  {NoStop}%
\bibitem [{\citenamefont {Sandvik}(1992)}]{P:sandvik92}%
  \BibitemOpen
  \bibfield  {author} {\bibinfo {author} {\bibfnamefont {A.~W.}\ \bibnamefont
  {Sandvik}},\ }\href {\doibase 10.1088/0305-4470/25/13/017} {\bibfield
  {journal} {\bibinfo  {journal} {J. Phys. A}\ }\textbf {\bibinfo {volume}
  {25}},\ \bibinfo {pages} {3667} (\bibinfo {year} {1992})}\BibitemShut
  {NoStop}%
\bibitem [{\citenamefont {Sylju\aa{}sen}\ and\ \citenamefont
  {Sandvik}(2002)}]{P:sandvik02}%
  \BibitemOpen
  \bibfield  {author} {\bibinfo {author} {\bibfnamefont {O.~F.}\ \bibnamefont
  {Sylju\aa{}sen}}\ and\ \bibinfo {author} {\bibfnamefont {A.~W.}\ \bibnamefont
  {Sandvik}},\ }\href {\doibase 10.1103/PhysRevE.66.046701} {\bibfield
  {journal} {\bibinfo  {journal} {Phys. Rev. E}\ }\textbf {\bibinfo {volume}
  {66}},\ \bibinfo {pages} {046701} (\bibinfo {year} {2002})}\BibitemShut
  {NoStop}%
\bibitem [{\citenamefont {Alet}\ \emph
  {et~al.}(2005{\natexlab{a}})\citenamefont {Alet}, \citenamefont {Wessel},\
  and\ \citenamefont {Troyer}}]{P:alet05}%
  \BibitemOpen
  \bibfield  {author} {\bibinfo {author} {\bibfnamefont {F.}~\bibnamefont
  {Alet}}, \bibinfo {author} {\bibfnamefont {S.}~\bibnamefont {Wessel}}, \ and\
  \bibinfo {author} {\bibfnamefont {M.}~\bibnamefont {Troyer}},\ }\href
  {\doibase 10.1103/PhysRevE.71.036706} {\bibfield  {journal} {\bibinfo
  {journal} {Phys. Rev. E}\ }\textbf {\bibinfo {volume} {71}},\ \bibinfo {eid}
  {036706} (\bibinfo {year} {2005}{\natexlab{a}})}\BibitemShut {NoStop}%
\bibitem [{\citenamefont {Alet}\ \emph
  {et~al.}(2005{\natexlab{b}})\citenamefont {Alet}, \citenamefont {Dayal},
  \citenamefont {Grzesik}, \citenamefont {Honecker}, \citenamefont {K\"orner},
  \citenamefont {L\"auchli}, \citenamefont {Manmana}, \citenamefont
  {McCulloch}, \citenamefont {Michel}, \citenamefont {Noack}, \citenamefont
  {Schmid}, \citenamefont {Schollw\"ock}, \citenamefont {St\"ockli},
  \citenamefont {Todo}, \citenamefont {Trebst}, \citenamefont {Troyer},
  \citenamefont {Werner},\ and\ \citenamefont {Wessel}}]{P:ALPS05}%
  \BibitemOpen
  \bibfield  {author} {\bibinfo {author} {\bibfnamefont {F.}~\bibnamefont
  {Alet}}, \bibinfo {author} {\bibfnamefont {P.}~\bibnamefont {Dayal}},
  \bibinfo {author} {\bibfnamefont {A.}~\bibnamefont {Grzesik}}, \bibinfo
  {author} {\bibfnamefont {A.}~\bibnamefont {Honecker}}, \bibinfo {author}
  {\bibfnamefont {M.}~\bibnamefont {K\"orner}}, \bibinfo {author}
  {\bibfnamefont {A.}~\bibnamefont {L\"auchli}}, \bibinfo {author}
  {\bibfnamefont {S.~R.}\ \bibnamefont {Manmana}}, \bibinfo {author}
  {\bibfnamefont {I.~P.}\ \bibnamefont {McCulloch}}, \bibinfo {author}
  {\bibfnamefont {F.}~\bibnamefont {Michel}}, \bibinfo {author} {\bibfnamefont
  {R.~M.}\ \bibnamefont {Noack}}, \bibinfo {author} {\bibfnamefont
  {G.}~\bibnamefont {Schmid}}, \bibinfo {author} {\bibfnamefont
  {U.}~\bibnamefont {Schollw\"ock}}, \bibinfo {author} {\bibfnamefont
  {F.}~\bibnamefont {St\"ockli}}, \bibinfo {author} {\bibfnamefont
  {S.}~\bibnamefont {Todo}}, \bibinfo {author} {\bibfnamefont {S.}~\bibnamefont
  {Trebst}}, \bibinfo {author} {\bibfnamefont {M.}~\bibnamefont {Troyer}},
  \bibinfo {author} {\bibfnamefont {P.}~\bibnamefont {Werner}}, \ and\ \bibinfo
  {author} {\bibfnamefont {S.}~\bibnamefont {Wessel}},\ }\href
  {http://jpsj.ipap.jp/link?JPSJS/74S/30/} {\bibfield  {journal} {\bibinfo
  {journal} {J. Phys. Soc. Jpn. S}\ }\textbf {\bibinfo {volume} {74}},\
  \bibinfo {eid} {30} (\bibinfo {year} {2005}{\natexlab{b}})}\BibitemShut
  {NoStop}%
\bibitem [{\citenamefont {Albuquerque}\ \emph {et~al.}(2007)\citenamefont
  {Albuquerque}, \citenamefont {Alet}, \citenamefont {Corboz}, \citenamefont
  {Dayal}, \citenamefont {Feiguin}, \citenamefont {Fuchs}, \citenamefont
  {Gamper}, \citenamefont {Gull}, \citenamefont {G\"urtler}, \citenamefont
  {Honecker}, \citenamefont {Igarashi}, \citenamefont {K\"orner}, \citenamefont
  {Kozhevnikov}, \citenamefont {L\"auchli}, \citenamefont {Manmana},
  \citenamefont {Matsumoto}, \citenamefont {McCulloch}, \citenamefont {Michel},
  \citenamefont {Noack}, \citenamefont {Paw{\l}owski}, \citenamefont {Pollet},
  \citenamefont {Pruschke}, \citenamefont {Schollw\"ock}, \citenamefont {Todo},
  \citenamefont {Trebst}, \citenamefont {Troyer}, \citenamefont {Werner},\ and\
  \citenamefont {Wessel}}]{P:ALPS07}%
  \BibitemOpen
  \bibfield  {author} {\bibinfo {author} {\bibfnamefont {A.~F.}\ \bibnamefont
  {Albuquerque}}, \bibinfo {author} {\bibfnamefont {F.}~\bibnamefont {Alet}},
  \bibinfo {author} {\bibfnamefont {P.}~\bibnamefont {Corboz}}, \bibinfo
  {author} {\bibfnamefont {P.}~\bibnamefont {Dayal}}, \bibinfo {author}
  {\bibfnamefont {A.}~\bibnamefont {Feiguin}}, \bibinfo {author} {\bibfnamefont
  {S.}~\bibnamefont {Fuchs}}, \bibinfo {author} {\bibfnamefont
  {L.}~\bibnamefont {Gamper}}, \bibinfo {author} {\bibfnamefont
  {E.}~\bibnamefont {Gull}}, \bibinfo {author} {\bibfnamefont {S.}~\bibnamefont
  {G\"urtler}}, \bibinfo {author} {\bibfnamefont {A.}~\bibnamefont {Honecker}},
  \bibinfo {author} {\bibfnamefont {R.}~\bibnamefont {Igarashi}}, \bibinfo
  {author} {\bibfnamefont {M.}~\bibnamefont {K\"orner}}, \bibinfo {author}
  {\bibfnamefont {A.}~\bibnamefont {Kozhevnikov}}, \bibinfo {author}
  {\bibfnamefont {A.}~\bibnamefont {L\"auchli}}, \bibinfo {author}
  {\bibfnamefont {S.}~\bibnamefont {Manmana}}, \bibinfo {author} {\bibfnamefont
  {M.}~\bibnamefont {Matsumoto}}, \bibinfo {author} {\bibfnamefont {I.~P.}\
  \bibnamefont {McCulloch}}, \bibinfo {author} {\bibfnamefont {F.}~\bibnamefont
  {Michel}}, \bibinfo {author} {\bibfnamefont {R.~M.}\ \bibnamefont {Noack}},
  \bibinfo {author} {\bibfnamefont {G.}~\bibnamefont {Paw{\l}owski}}, \bibinfo
  {author} {\bibfnamefont {L.}~\bibnamefont {Pollet}}, \bibinfo {author}
  {\bibfnamefont {T.}~\bibnamefont {Pruschke}}, \bibinfo {author}
  {\bibfnamefont {U.}~\bibnamefont {Schollw\"ock}}, \bibinfo {author}
  {\bibfnamefont {S.}~\bibnamefont {Todo}}, \bibinfo {author} {\bibfnamefont
  {S.}~\bibnamefont {Trebst}}, \bibinfo {author} {\bibfnamefont
  {M.}~\bibnamefont {Troyer}}, \bibinfo {author} {\bibfnamefont
  {P.}~\bibnamefont {Werner}}, \ and\ \bibinfo {author} {\bibfnamefont
  {S.}~\bibnamefont {Wessel}},\ }\href {\doibase 10.1016/j.jmmm.2006.10.304}
  {\bibfield  {journal} {\bibinfo  {journal} {J. Magn. Magn. Mat.}\ }\textbf
  {\bibinfo {volume} {310}},\ \bibinfo {eid} {1187} (\bibinfo {year}
  {2007})}\BibitemShut {NoStop}%
\bibitem [{\citenamefont {Hukushima}\ and\ \citenamefont
  {Nemoto}(1996)}]{P:hukushima96}%
  \BibitemOpen
  \bibfield  {author} {\bibinfo {author} {\bibfnamefont {K.}~\bibnamefont
  {Hukushima}}\ and\ \bibinfo {author} {\bibfnamefont {K.}~\bibnamefont
  {Nemoto}},\ }\href {\doibase 10.1143/JPSJ.65.1604} {\bibfield  {journal}
  {\bibinfo  {journal} {J. Phys. Soc. Jpn.}\ }\textbf {\bibinfo {volume}
  {65}},\ \bibinfo {pages} {1604} (\bibinfo {year} {1996})}\BibitemShut
  {NoStop}%
\bibitem [{\citenamefont {Katzgraber}\ \emph {et~al.}(2006)\citenamefont
  {Katzgraber}, \citenamefont {Trebst}, \citenamefont {Huse},\ and\
  \citenamefont {Troyer}}]{P:katzgraber06}%
  \BibitemOpen
  \bibfield  {author} {\bibinfo {author} {\bibfnamefont {H.~G.}\ \bibnamefont
  {Katzgraber}}, \bibinfo {author} {\bibfnamefont {S.}~\bibnamefont {Trebst}},
  \bibinfo {author} {\bibfnamefont {D.~A.}\ \bibnamefont {Huse}}, \ and\
  \bibinfo {author} {\bibfnamefont {M.}~\bibnamefont {Troyer}},\ }\href
  {\doibase 10.1088/1742-5468/2006/03/P03018} {\bibfield  {journal} {\bibinfo
  {journal} {J. Stat. Mech.: Theory and Experiment}\ ,\ \bibinfo {pages}
  {P03018}} (\bibinfo {year} {2006})}\BibitemShut {NoStop}%
\bibitem [{\citenamefont {Melko}(2007)}]{P:melko07}%
  \BibitemOpen
  \bibfield  {author} {\bibinfo {author} {\bibfnamefont {R.~G.}\ \bibnamefont
  {Melko}},\ }\href {\doibase 10.1088/0953-8984/19/14/145203} {\bibfield
  {journal} {\bibinfo  {journal} {J. Phys.: Cond. Matt.}\ }\textbf {\bibinfo
  {volume} {19}},\ \bibinfo {pages} {145203} (\bibinfo {year}
  {2007})}\BibitemShut {NoStop}%
\bibitem [{\citenamefont {Binder}(1981{\natexlab{a}})}]{P:binder81L}%
  \BibitemOpen
  \bibfield  {author} {\bibinfo {author} {\bibfnamefont {K.}~\bibnamefont
  {Binder}},\ }\href {\doibase 10.1103/PhysRevLett.47.693} {\bibfield
  {journal} {\bibinfo  {journal} {Phys. Rev. Lett.}\ }\textbf {\bibinfo
  {volume} {47}},\ \bibinfo {pages} {693} (\bibinfo {year}
  {1981}{\natexlab{a}})}\BibitemShut {NoStop}%
\bibitem [{\citenamefont {Binder}(1981{\natexlab{b}})}]{P:binder81Z}%
  \BibitemOpen
  \bibfield  {author} {\bibinfo {author} {\bibfnamefont {K.}~\bibnamefont
  {Binder}},\ }\href {\doibase 10.1007/BF01293604} {\bibfield  {journal}
  {\bibinfo  {journal} {Z. Phys. B}\ }\textbf {\bibinfo {volume} {43}},\
  \bibinfo {pages} {119} (\bibinfo {year} {1981}{\natexlab{b}})}\BibitemShut
  {NoStop}%
\bibitem [{\citenamefont {Pollock}\ and\ \citenamefont
  {Ceperley}(1987)}]{P:pollock87}%
  \BibitemOpen
  \bibfield  {author} {\bibinfo {author} {\bibfnamefont {E.~L.}\ \bibnamefont
  {Pollock}}\ and\ \bibinfo {author} {\bibfnamefont {D.~M.}\ \bibnamefont
  {Ceperley}},\ }\href {\doibase 10.1103/PhysRevB.36.8343} {\bibfield
  {journal} {\bibinfo  {journal} {Phys. Rev. B}\ }\textbf {\bibinfo {volume}
  {36}},\ \bibinfo {pages} {8343} (\bibinfo {year} {1987})}\BibitemShut
  {NoStop}%
\bibitem [{\citenamefont {Misguich}\ \emph {et~al.}(2002)\citenamefont
  {Misguich}, \citenamefont {Lhuillier}, \citenamefont {Mambrini},\ and\
  \citenamefont {Sindzingre}}]{P:misguich02}%
  \BibitemOpen
  \bibfield  {author} {\bibinfo {author} {\bibfnamefont {G.}~\bibnamefont
  {Misguich}}, \bibinfo {author} {\bibfnamefont {C.}~\bibnamefont {Lhuillier}},
  \bibinfo {author} {\bibfnamefont {M.}~\bibnamefont {Mambrini}}, \ and\
  \bibinfo {author} {\bibfnamefont {P.}~\bibnamefont {Sindzingre}},\ }\href
  {\doibase 10.1140/epjb/e20020078} {\bibfield  {journal} {\bibinfo  {journal}
  {Eur. Phys. J. B}\ }\textbf {\bibinfo {volume} {26}},\ \bibinfo {pages} {167}
  (\bibinfo {year} {2002})}\BibitemShut {NoStop}%
\bibitem [{Note1()}]{Note1}%
  \BibitemOpen
  \bibinfo {note} {The code for the \protect \emph {spinpack} is available at
  http://www.ovgu.de/jschulen/spin}\BibitemShut {NoStop}%
\bibitem [{\citenamefont {Majumdar}(2010)}]{P:majumdar10}%
  \BibitemOpen
  \bibfield  {author} {\bibinfo {author} {\bibfnamefont {K.}~\bibnamefont
  {Majumdar}},\ }\href {\doibase 10.1103/PhysRevB.82.144407} {\bibfield
  {journal} {\bibinfo  {journal} {Phys. Rev. B}\ }\textbf {\bibinfo {volume}
  {82}},\ \bibinfo {pages} {144407} (\bibinfo {year} {2010})}\BibitemShut
  {NoStop}%
\bibitem [{\citenamefont {Sandvik}\ \emph {et~al.}(2002)\citenamefont
  {Sandvik}, \citenamefont {Daul}, \citenamefont {Singh},\ and\ \citenamefont
  {Scalapino}}]{P:sandvik02b}%
  \BibitemOpen
  \bibfield  {author} {\bibinfo {author} {\bibfnamefont {A.~W.}\ \bibnamefont
  {Sandvik}}, \bibinfo {author} {\bibfnamefont {S.}~\bibnamefont {Daul}},
  \bibinfo {author} {\bibfnamefont {R.~R.~P.}\ \bibnamefont {Singh}}, \ and\
  \bibinfo {author} {\bibfnamefont {D.~J.}\ \bibnamefont {Scalapino}},\ }\href
  {\doibase 10.1103/PhysRevLett.89.247201} {\bibfield  {journal} {\bibinfo
  {journal} {Phys. Rev. Lett.}\ }\textbf {\bibinfo {volume} {89}},\ \bibinfo
  {pages} {247201} (\bibinfo {year} {2002})}\BibitemShut {NoStop}%
\end{thebibliography}%

\end{document}